\documentclass[twoside]{article}
\def\R{{I\!\! R}}
\def\N{{I\!\! N}}

\def\C{~\hbox{\vrule width 0.6pt height 6pt depth 0pt \hskip -3.5pt}C}
\def\kasten{$~~\mbox{\hfil\vrule height6pt width5pt depth-1pt}$ }

\newtheorem{theorem}{Theorem}[section]
\newtheorem{definition}[theorem]{Definition}
\newtheorem{proposition}[theorem]{Proposition}
\newtheorem{condition}[theorem]{Condition}
\newtheorem{corollary}[theorem]{Corollary}
\title{SPDEs leading to local, relativistic
 quantum vector fields with indefinite metric and nontrivial S-matrix}
\author{Sergio Albeverio and Hanno Gottschalk\\ Institut f\"ur angewandte Mathematik\\
Rheinische Friedrich-Wilhelms-Universit\"at Bonn\\
D-53115 Bonn, Germany\\ albeverio@uni-bonn.de/gottscha@wiener.iam.uni-bonn.de\\ \\
Jiang-Lun Wu \\ Department of Mathematics, University of Wales Swansea\\ Singelton Park, Swansea SA2 8PP\\
Wales, UK\\ 
j.l.wu@swansea.ac.uk}

\begin{document}
\maketitle
\pagestyle{myheadings}
\thispagestyle{empty}
\markboth{S. Albeverio, H. Gottschalk, J.-L. Wu}{SPDEs and QFT with indefinite metric}
\begin{abstract}
In this article we review the construction of local, relativistic
quantum vector fields by analytic continuation of Euclidean vector
fields obtained as solutions of covariant SPDEs. We revise the formulation
of such SPDEs by introducing new Gaussian noise terms -- a procedure which avoids
the re-definition of the two point functions needed in previous
approaches in order to obtain relativistic fields with nontrivial scattering. 
We describe the construction of asymptotic states and the scattering of the
analytically continued solutions of these new SPDEs and we give precise 
conditions for nontrivial and well-defined scattering.   
\end{abstract}
{\small \noindent{\bf MSC (2000):} \underline{81T08}, 60H15, 60G20

\noindent{\bf Key Words:} {\it SPDEs driven by non-Gaussian white noise; relativistic Quantum vector fields with indefinite metric.}}

\section{Introduction}
Since the work of Symanzik and Nelson \cite{Ne1,Ne2,Sy}, the construction of local,
relativistic quantum fields via analytic continuation from Euclidean random fields
has been the most vital and productive paradigm in constructive quantum field theory
(QFT), see e.g. the by now classical expositions in \cite{GJ,Si}. This Euclidean strategy
has been completed successfully in $d=2$ space-time dimensions and partial results
have been obtained for $d=3$ \cite{Ba,GJ0}. In the physical space-time dimension $d=4$ however,
the standard approach to the definition of local potentials via renormalization up to
now is plagued by seemingly incurable ultra-violet divergences, and no construction of
a non-trivial (interacting) quantum field is known within that approach.

In a series of papers beginning with the mid-80ies \cite{AH1,AH2,AH3}, a different approach was
suggested for $d=4$ dimensions, where Euclidean random fields are being constructed as
solutions of the quaternionic Cauchy-Riemann partial differential equation 
driven by a non Gaussian white noise. For the so-defined models, the analytic continuation
to relativistic fields of self-interacting electro-magnetic type can be performed \cite{AIK}
and the corresponding relativistic models have non-trivial scattering behavior \cite{AGW3}. The
physical interpretation of the given fields, i.e. a solution of the Gauge problem in the
sense of Gupta and Bleuler \cite{Gu,Bl} (i.e. the identification of the physical Hilbert space), is still an unsolved problem and therefore these models
can not yet be considered as yielding fully satisfactory 4-d relativistic quantum fields. We however note 
that one can verify the modified Wightman axioms for quantum fields with indefinite metric \cite{MS,Str}
and  the asymptotic incoming and outgoing fields can be gauged as in usual free quantum 
electrodynamics. Furthermore, gauge invariance of the scattering (S-) matrix has been verified in
 \cite{AGW3}.

In \cite{BGL,GL1,GL2} a generalization of this program was suggested by Becker, Gielerak and \L ugiewicz and the equation $DA=F$ was
studied systematically in arbitrary space-time dimension, where $D$ is a covariant partial
differential operator with constant coefficients and $F$ is a generalized (non-Gaussian)
white noise vector field. Solution of this SPDE under suitable conditions on $D$ were given in
\cite{BGL}, vacuum expectation values of the associated relativistic fields are given in
\cite{AGW4,BGL,GL1,GL2} and the scattering behavior has been calculated in \cite{AGW4}.

Here we review the construction of relativistic quantum vector fields with indefinite metric
via solutions of $DA=F$. Most of the material in this article has ben taken from the
above references, but we introduce some technical improvements in the construction of the
vector noise $F$ by adding new Gaussian noise terms (depending on $D$) which render obsolete
the re-definition of relativistic two point functions in \cite{AGW4}. This re-definition in \cite{AGW4}
was necessary to obtain a well-defined scattering behavior for a certain class of models, but it unfortunately
destroyed the direct connection of quantum fields with the random field models. We here demonstrate that a modification
of the SPDE can do the same job and the relativistic and Euclidean side remain directly connected.

The article contains the following materials: In Section 2 we introduce the generalized vector noise
fields, we discuss covariance properties of the operator $D$ and we set up and solve a modified SPDE
$DA=F+F^g$. We calculate two- and $n$-point Schwinger functions (moments) of the random field
$A$ and we show that under suitable conditions on $D$ the two point function of $A$ is a superposition of
two point functions of free Euclidean vector fields. Section 3 deals with the analytic continuation of the 
Schwinger functions to relativistic Wightman functions. Section 4 is devoted to the construction of quantum fields with indefinite metric s.t. the Wightman functions
constructed in Section 3 are the vacuum expectation values of these fields. Scattering
of the relativistic fields is being discussed in section 5: Under given conditions on $D$ ("no dipole condition", cf. \cite{Go2} for a physical interpretation),
we construct asymptotic states and explicitly calculate the scattering ($S$-) matrix of the models following \cite{AG}.
         
\section{Construction of Euclidean random fields via (modified) covariant SPDEs}

Let $d$ be the space-time dimension, we then identify the Euclidean space-time with $\R^d$.
Let $\tau:{\rm SO}(d)\to {\rm Gl}(L)$ be a $L$-dimensional (not necessarily irreducible)
real representation of the orthogonal group. Since ${\rm SO}(d)$ is compact, we may assume without
loss of generality that $\tau:{\rm SO}(d)\to{\rm SO}(L)$. We want to study covariant noise vector fields over $\R^d$,
which are defined as follows: 

A (tempered) \underline{random vector field} of dimension $L$ over $\R^d$ is a mapping $\phi$
 from the space of $\R^L$-valued Schwarz test functions ${\cal S}={\cal S}(\R^d,\R^L)$ to the
real valued random variables of some probability space $(\Omega,
{\cal B},P)$ such that 1) $\phi$ is linear $P$ a.s. and 2) $f_n\to f$ in ${\cal S}$
$\Rightarrow$ $\phi(f_n)\to\phi(f)$ in probability law. $\phi$ by definition \underline{transforms
covariantly under $\tau$ } if $\phi(f)=\phi(f_g)$ in probability law $\forall f\in{\cal S}$ where $g\in{\rm SO}(d)$ and 
$f_g(x)=\tau(g)f(g^{-1}x)$. Furthermore, $\phi$ is called \underline{translation invariant} or \underline{stationary}
if $\phi(f_a)=\phi(f)$ in law $\forall f\in{\cal S}$ where $a\in\R^d$ and $f_a(x)=f(x-a)$.

\begin{definition}
\label{2.1def}
 A $\tau$-covariant noise field is an $L$-dimensional random vector
field over $\R^d$ which transforms covariantly under $\tau$, is stationary and 
is independent when localized in non intersecting regions (i.e. $\phi(f)$ is independent of
$\phi(h)$ if ${\rm supp}f\cap{\rm supp} h=\emptyset$.)
\end{definition}

A \underline{characteristic functional} is a mapping ${\cal C}:{\cal S}\to\C$ fulfilling
1) ${\cal C}$ is continuous, 2) $\cal C$ is positive definite and 3) $\cal C$ is normalized 
${\cal C}(0)=1$. By Minlos' theorem \cite{I,Mi} there is a one-to-one (up to equivalence in law) 
correspondence between ($L$-dimensional) random fields $\phi$ and characteristic functionals
 ${\cal C}_\phi$ given by ${\cal C}_\phi(f)=E[e^{i\phi(f)}]$. Furthermore, $\phi$ can be realized
 as coordinate process on the measurable space $({\cal S}',{\cal B})$, with ${\cal S}'$
 the topological dual space of $\cal S$ and ${\cal B}$ the Borel sigma algebra on ${\cal S}'$. Explicitly,
 given a random vector field $\phi$ there exists a unique probability measure $P_\phi$ on this
  measurable space such that $\phi(f)=\langle .,f\rangle~\forall f\in{\cal S}$ in law where $\langle.,.\rangle$ is the
  dualization of ${\cal S}'$ and $\cal S$. Thus the path space of $\phi$ is always contained in 
  the space of tempered distributions.     

As a simple consequence of Minlos' theorem we get that $F$ is a $\tau$-covariant noise if and only if
the associated characteristic functional ${\cal C}_F$ fulfills ${\cal C}_F(f_g)=
{\cal C}_F(f_a)={\cal C}_F(f)\forall f\in{\cal S},g\in{\rm SO}(d),a\in\R^d$ and ${\cal C}_F
(f+h)={\cal C}_F(f){\cal C}_F(h)\forall f,h\in{\cal S}$ s.t. ${\rm supp}f\cap{\rm supp}h=\emptyset$.
To construct noise fields, it is thus sufficient to define a characteristic functional with
the above properties.

Let $\psi$ be a $C^\infty$ L\'evy characteristic over $\R^L$, i.e.
\begin{equation}
\label{2.1eqa}
\psi(t)=ia\cdot t-{\sigma t\cdot\sigma t\over 2}+\int_{\R^L-\{ 0\}}\left[e^{it\cdot s}-1\right]dM(s)
\end{equation}
where $t,a\in\R^L,\sigma\in{\rm Mat_{L\times L}(\R)}$ is symmetric and positive semi definite
and the L\'evy measure $M$ is a measure on $\R^L-\{0\}$ s.t. $M$ has moments
of all orders. $\psi$ is uniquely determined by $a$ (deterministic part), $\sigma$ (Gaussian part) and $M$ (Poisson part).
We say that $\psi$ is $\tau$-invariant if $\psi(\tau(g)t)=\psi(t)\forall g\in{\rm SO}(\R^d), t\in\R^L$.
Obviously, this is the case if and only if $\tau(g)a=a$, $\sigma^2$ commutes with $\tau(g)$ and
$\tau(g)_*M=M$ $\forall g\in{\rm SO}(d)$. 

\begin{proposition}
\label{2.1prop}
Let $\psi$ be a $C^\infty$ and $\tau$-invariant L\'evy characteristic. Then (i)
\begin{equation}
\label{2.2eqa}
{\cal C}_F(f)=\exp\left[\int_{\R^d}\psi(f)\,dx\right],~~~f\in{\cal S}
\end{equation}
is a characteristic functional. The associated random vector field $F$ is a $\tau$-covariant
 noise field.
 
\noindent (ii) Let $\Delta$ be the Laplacian on $\R^d$,
$p:\R\to\R$ be a polynomial which is positive semi-definite on $[0,\infty)$ and
$\bar\sigma^2=-\left({\partial^2\psi(t)\over \partial t_\alpha \partial t_\beta}|_{t=0}\right)_{\alpha,\beta=1,\ldots,L}$. Then
\begin{equation}
\label{2.3eqa}
{\cal C}_{F^g}(f)=\exp\left[ -\int_{\R^d} \bar\sigma f\cdot \bar\sigma p(-\Delta)f\, dx  \right],~~~ f\in{\cal S}  
\end{equation}
defines a Gaussian $\tau$-covariant noise field. 
 \end{proposition}
 \noindent {\bf Proof.} (i) Continuity of ${\cal C}_F$ follows from $\psi$ being $C^\infty$,
 normalization is a consequence of $\psi(0)=0$ and positive definiteness can be derived
 from the fact that $\psi$ is a conditionally positive definite\footnote{$\sum_{l,j=1}^n
 \psi(t_l-t_j)z_l\bar z_j\geq 0$ if $\sum_{l=1}^n z_j=0$, $t_j\in\R^L,z_j\in\C$, $j=1,\ldots,n; n\in\N$}
 function in $t$ and thus  $\int_{\R^d}\psi(f)\, dx$ is conditionally positive definite
 in $f$. But the exponential of a conditionally positive definite function is positive definite
 by Schoenberg's theorem \cite{BF}. $\tau$-invariance of ${\cal C}_F$ follows from the
 $\tau$-invariance of $\psi$ and the invariance of the Lebesgue measure under orthogonal transformations.
 Likewise, invariance of ${\cal C}_F$ under translations follows from translation 
 invariance of $dx$. That ${\cal C}_F(f+h)$ factors for ${\rm supp} f\cap {\rm supp} h=\emptyset$
 is implied by $\psi(0)=0$. 
 
 (ii) The proof that ${\cal C}_{F^g}$ is the characteristic functional of a 
 Gaussian random field is standard (we note that $\bar\sigma$ by definition is 
 positive semi-definite and commutes with the positive operator $p(-\Delta)$.) Invariance follows from
 the fact that $\bar \sigma^2$ commutes with $\tau(g)$ and $p(-\Delta)$ is invariant 
 under orthogonal transformations. Translation invariance follows as in (i). That
 ${\cal C}_{F^g}(f+h)$ factors if ${\rm supp}f\cap{\rm supp}h=\emptyset$ is a 
 consequence of
 $$
 \int_{\R^d} \bar\sigma f\cdot \bar\sigma p(-\Delta)h\, dx=0
 $$
 for $f,h$ as above. \kasten 

 Let $D:{\cal S}\to{\cal S}$ be a $\tau$-covariant partial differential operator with constant
 coefficients, i.e. $Df_g=(Df)_g$ and $Df_a=(Df)_a$ $\forall f\in{\cal S}, g\in{\rm SO}(d),a\in\R^d$.
 Assuming that $D$is continuously invertible, the following representation has been obtained \cite{BGL} for
 the Fourier transform of the Green's function of $D$:
\begin{equation}
\label{2.4eqa}
\hat D^{-1}(k)= {Q_E(k)\over \prod_{l=1}^N(|k|^2+m_l^2)^{\nu_l}}
\end{equation}
with $m_l\in\C-(-\infty,0],m_j\not=m_l$ for $ l\not = j$ and $\nu_l\in \N$.
$Q_E(k)$ is an $L\times L$-matrix with polynomial entries of order
$\leq \kappa=2(\sum_{l=1}^N\nu_l-1)$ which fulfills the Euclidean
transformation law ${\rm ad }_{\tau(g)}( Q_E(g^{-1}k))=Q_E(k)\forall
g\in {\rm SO}(d)$. Without loss of generality we assume that
$Q_E$ is prime w.r.t the factors $(|k|^2+m_l^2)$, i.e. that none of
them divides all of the polynomial matrix elements of $Q_E$ and furthermore
we impose  a "positive mass spectrum" condition $m_l\in(0,\infty)$ for
$l=1,\ldots,N$.

Given this representation of $\hat D^{-1}(k)$ we define
\begin{equation}
\label{2.5eqa}
p(t)=p(t,D)={\prod_{l=1}^N(t+m_l^2)^{\nu_l}\over \prod_{l=1}^N m_l^{2\nu_l}}-1
\end{equation}
and we note that $p(t)\geq 0$ as $t\geq 0$.

Let $\psi$ be a $\tau$-invariant L\'evy characteristic s.t. $d\psi (t)/dt_\beta|_{t=0} =0$ and $D$ as above. We define $F$ as in
Proposition \ref{2.1prop} (i) and $F^g$ as in Proposition \ref{2.1prop} (ii) with $p(t)=p(t,D)$ as in
Equation (\ref{2.5eqa}), $F^g$ independent of $F$ (in the stochstic sense). We set up the crucial SPDE of this work as
\begin{equation}
\label{2.6eqa}
DA=F+F^g
\end{equation}
which can be solved pathwisely in ${\cal S}'$ since the random field $F+F^g$ by
Minlos' theorem has paths in ${\cal S}'$ and $D$ by our assumptions (and duality) is continuously
invertible on that space. We get for the properties of $A$:

\begin{theorem}
\label{2.1theo}
$A$ obtained as the unique solution of (\ref{2.6eqa}) in ${\cal S}'$ is a $\tau$-covariant random vector field. The Schwinger
functions (moments) of $A$ are given by
\begin{eqnarray}
\label{2.9eqa}
S_{n,\alpha_1\cdots\alpha_n}(x_1,\ldots,x_n)&=& E\left[A_{\alpha_1}(x_1)\cdots A_{\alpha_n}(x_n)\right]\nonumber \\
&=&\sum_{I\in{\cal P}^{(n)}}\prod_{\{j_1,\ldots,j_l\}\in I} S_{l,\alpha_{j_1}\cdots\alpha_{j_l}}^T(x_{j_1},\ldots,x_{j_l})
\end{eqnarray}
where ${\cal P}^{(n)}$ is the collection of all partitions of $\{1,\ldots,n\}$ into nonempty subsets.
For $n=2$ 
\begin{equation}
\label{2.13eqa}
S_{2,\alpha_1\alpha_2}^T(x_1,x_2)={Q^E_{2,\alpha_1\alpha_2}(-i\underline{\nabla}_2)\over \prod_{l=1}^Nm_l^{2\nu_l}}\left[\prod_{l=1}^N(-\Delta+m_l^2)^{-\nu_l}\right](x_1-x_2)
\end{equation}
and for $n\geq 3$
\begin{eqnarray}
\label{2.14eqa}
S^T_{n,\alpha_1\cdots\alpha_n}(x_1,\ldots,x_n) &=& Q^E_{n,\alpha_1\cdots\alpha_n}(-i\underline{\nabla}_n)\nonumber\\
&\times&\int_{\R^d}\prod_{j=1}^n\left[\prod_{l=1}^N(-\Delta+m_l^2)^{-\nu_l}\right](x_j-x) \, dx
\end{eqnarray}
where
\begin{equation}
\label{2.11eqa}
Q^E_{n,\alpha_1\cdots\alpha_n}(-i\underline{\nabla}_n)=
C^{\beta_1\cdots\beta_n}\prod_{l=1}^nQ_{E,\beta_l,\alpha_l}(-i{\partial\over\partial x_l})
\end{equation}
with
\begin{equation}
\label{2.12eqa}
C_{\beta_1\cdots\beta_n}=(-i)^n\left.{\partial^n\psi(t)\over \partial t_{\beta_1}\cdots \partial t_{\beta_n}}\right|_{t=0}
\end{equation}
and we applied the Einstein convention of summation and uppering/lowering of indices on 
$\R^L$ w.r.t. the invariant inner product $\cdot$. The Schwinger functions fulfill the requirements of $\tau$-covariance,
translation invariance, symmetry, clustering and Hermiticity from the Osterwalder-Schrader
axioms \cite{OS1,OS2} of Euclidean QFT.
\end{theorem}
\noindent {\bf Proof.} We give a short outline of the proof: $\tau$-covariance of $A$ and stationarity can be deduced from the $\tau$-
 (and translation-)invariance of the characteristic functional ${\cal C}_A(f)={\cal C}_F(D^{-1}f){\cal C}_{F^g}(D^{-1}f)$.
 This property follows from the related property for ${\cal C}_F$ and ${\cal C}_{F^g}$ and $\tau$-covariance (translation invariance)
 of $D$.
The characteristic functional ${\cal C}_A$ is given by the following explicit formula
\begin{equation}
\label{2.7eqa}
{\cal C}_A(f)=\exp\left[\int_{\R^d}\psi(D^{-1}f)-\bar\sigma D^{-1}f\cdot \bar\sigma p(-\Delta)D^{-1}f\,dx\right]~~~f\in{\cal S}
\end{equation}
and we can calculate the Schwinger functions (or moments) of $A$ by
\begin{eqnarray}
\label{2.8eqa}
S_{n,\alpha_1\cdots\alpha_n}(x_1,\ldots,x_n)
= \left.(-i)^n{\delta^n\over\delta A_{\alpha_1}(x_1)\cdots \delta A_{\alpha_n}(x_n)}{\cal C}_A(f)\right|_{f=0}.
\end{eqnarray}
Using the linked cluster identity (see e.g. \cite{AGW1}) we get
\begin{equation}
\label{2.10eqa}
S_{n,\alpha_1\cdots\alpha_n}^T(x_1,\ldots,x_n)=
 \left.(-i)^n{\delta^n\over\delta A_{\alpha_1}(x_1)\cdots \delta A_{\alpha_n}(x_n)}\ln{\cal C}_A(f)\right|_{f=0}.
\end{equation}
By Eqs. (\ref{2.4eqa})-(\ref{2.5eqa}),(\ref{2.7eqa}) and (\ref{2.10eqa}) one derives Eq. (\ref{2.13eqa}) and (\ref{2.14eqa})
by explicit calculations.

The $\tau$-covariance / translation invariance properties of the Schwinger functions follow from the
related properties of the random vector field $A$. Symmetry and Hermiticity are trivial (note that $A$ is a 
real vector field). The cluster property can be verified from the fact that the Green's function 
$$\left[\prod_{l=1}^N(-\Delta+m_l^2)^{-\nu_l}\right](x)$$ 
and its derivatives
are of exponential decay as $|x|\to\infty$, see \cite{AGW1} for a related situation.\kasten  

Here we would like to point out that the difference between $A$ as defined here and the corresponding fields 
defined in \cite{AGW4} is that
in that reference 
\begin{equation}
\label{2.15eqa}
S^T_{2\alpha_1,\alpha_2}(x_1-x_2)=Q^E_{2,\alpha_1\alpha_2}(-i\underline{\nabla}_2)\left[
\prod_{l=1}^N(-\Delta+m_l^2)^{-2\nu_l}\right](x_1-x_2)
\end{equation}
and the effect of the Gaussian noise $F^g$ is to replace (\ref{2.15eqa}) by (\ref{2.13eqa}). As we
shall see in the discussion of relativistic scattering theory below, this correction leads to a well-defined
particle like asymptotics, for $\nu_l=1,l=1,\ldots,n$, of the related relativistic quantum field models. Without this correction
(or the somewhat {\it ad hoc} replacement in \cite{AGW4}) the asymptotics would be dipole-like rather than particle like. Such effects
now only occur when some of the $\nu_l$ are strictly larger than one, cf. \cite{Go2}.

\begin{condition}
\label{2.1cond}
We say that the partial differential operator $D$ fulfills the \underline{no di-}\linebreak\underline{pole condition}, if in the representation (\ref{2.4eqa}) we have
$\nu_l=1$ for $l=1,\ldots,N$. We restrict from now on our models to fulfill this condition (cf. \cite{Go2} for the meaning
of this condition).
\end{condition}

\section{Analytic continuaton of the Schwinger functions}

In this section we discuss the analytic continuation of the truncated Schwinger
functions $S_n^T$ to relativistic truncated Wightman functions $\hat W_n^T$. A solution to this
problem can be obtained by representing $S_n^T$ as Fourier-Laplace transform, i.e.
\begin{eqnarray}
\label{3.1eqa}
S^T_{n,\alpha_1\cdots\alpha_n}(x_1,\ldots,x_n)
&=& (2\pi)^{-dn/2}\int_{\R^{dn}} \exp(\sum_{l=1}^n-k^0_l
x_l^0+i\vec k_l\cdot \vec x_l)\nonumber\\
&\times& \hat W_{n,\alpha_1\cdots\alpha_n}^T(k_1,\ldots,k_l)
~dk_1\cdots dk_n~,
\end{eqnarray}
where $x_1^0<\ldots<x_n^0$ and $\hat W_{n\alpha_1\cdots\alpha_n}^T$ ( the Fourier transform of $W_n^T$)
is a tempered distribution which fulfills the spectral property, i.e. it has support in the cone
$\{(k_1,\ldots,k_n)\in \R^{dn}:
q_j=\sum_{l=1}^j k_l \in \bar V_0^-, j=1,\ldots,n-1\}$.
 Here, $\bar V_0^-$ stands for the closed
backward lightcone (that
we do not use the forward lightcone for the formulation of the
spectral condition as done in some other references,
is a matter of convention on the Fourier transform).
Under this condition the above integral representation exists. From the general theory of quantum fields
it follows that $S_n^T$
is the analytic continuation $W_n^T$ from points with
purely relativistically real time to the Euclidean points of purely
imaginary time. Furthermore, it follows from the symmetry and Euclidean
covariance of the $S_n^T$ that $W_n^T$ fulfills the requirements of
Poincar\'e covariance  (w.r.t. the analytic continuation of the representation $\tau$)
and locality, see e.g. \cite{OS1}.

Following an idea of \cite{BGL}, we expand the denominator of Eq. (\ref{2.4eqa}) into partial fractions
(we recall that we assume Condition \ref{2.1cond} to hold)
\begin{equation}
\label{3.2eqa}
{1\over
\prod_{l=1}^N(|k|^2+m_l^2)}=\sum_{l=1}^N
{b_{l}\over (|k|^2+m_l^2)}
\end{equation}
with $b_{l}\in \R$ uniquely determined and $b_{l}\not =0$. Thus,
$D^{-1}(x)$ can be represented as $Q_E(-i\partial/\partial x) \sum_{l=1}^Nb_{l}(-\Delta+m_l^2)^{-1}(x)$ and for $n\geq 3$
\begin{eqnarray}
\label{3.3eqa}
S_{n,\alpha_1\cdots\alpha_n}^T(x_1,\ldots,x_n)&=&  Q_{n,\alpha_1
\cdots\alpha_n}^E(-i\underline{\nabla}_n)\sum_{l_1,\ldots,l_n=1}^N
\nonumber\\
&\times& \prod_{r=1}^nb_{l_r}\int_{\R^d}\prod _{j=1}^n(-\Delta+m_{l_j}^2)^{-1}(x-x_j)~dx
\end{eqnarray}
Setting
\begin{equation}
\label{3.4eqa}
S^T_{n,m_{1},\cdots
m_{n}}(x_1,\ldots,x_n)=\int_{\R^d}\prod_{j=1}^n(-
\Delta+m_{j}^2)^{-1}(x_j-x)~dx
\end{equation}
we note that a Fourier-Laplace representation (\ref{3.1eqa})of $S_{n,m_1,\ldots,m_n}^T(x_1,\ldots,x_n)$ has been calculated in
\cite{AGW1} Proposition 7.8. $\hat W_{n,m_1,\ldots,m_n}^T(k_1,\ldots,k_n)$ is given by
\begin{equation}
\label{3.5eqa}
(2\pi)^{-(d(n-2)-2)/2}\left\{\sum_{j=1}^n\prod_{l=1}^{j-1}\delta^-_{m_l}(k_l){(-1)\over k^2-m_j^2}\prod_{l=j+1}^n\delta^+_{m_l}(k_l)\right\}\delta(\sum_{l=1}^nk_l)
\end{equation}
Here $\delta_m^{\pm}(k)=\theta (\pm k^0)\delta(k^2-m^2)$ where $\theta$ is the Heaviside step function and $k^2={k^0}^2-|\vec
k|^2$. Furthermore, let
\begin{equation}
\label{3.5aeqa}
Q^M_{n}((k_1^0,\vec k_1),\ldots,
 (k_n^0,\vec k_n))=Q^E_{n}((ik_1^0,\vec k_1),\ldots,(i k_n^0,
\vec k_n)).
\end{equation}
We then define
\begin{equation}
\label{3.6eqa}
\hat W_{2,\alpha_1\alpha_2}^T(k_1,k_2)=(2\pi)^{(d+1)}{Q^M_{2,\alpha_1\alpha_2}(k_1,k_2)\over
\prod_{l=1}^Nm_l^2}\sum_{l=1}^Nb_l\,\delta_{m_l}^-(k_1)\delta(k_1+k_2)
\end{equation}
and
\begin{eqnarray}
\label{3.7eqa}
\hat W_{n,\alpha_1\cdots\alpha_n}^T(k_1,\ldots,k_n)&=&Q^M_{n,\alpha_1\cdots\alpha_n}( k_1,\ldots,
 k_n)\nonumber\\
 &\times&\sum_{l_1,\ldots,l_n=1}^N\prod_{j=1}^nb_{l_j} \hat W_{n,m_{l_1},\ldots,m_{l_n}}^T(k_1,\ldots,k_n)
\end{eqnarray}
 and we can now put together these pieces in the following theorem:

\begin{theorem}
\label{3.1theo}
The truncated Schwinger functions $S_n^T$ have a Fourier-Laplace representation (\ref{3.1eqa}) with
$\hat W_n^T$ defined in Eqs. (\ref{3.6eqa}) and (\ref{3.7eqa}).  Equivalently, $S_n^T$ is the analytic continuation of
$W_n^T$ from purely real relativistic time to purely imaginary Euclidean time.
The truncated Wightman functions $W_n^T$ fulfill the requirements of
temperedness, relativistic covariance w.r.t. the representation of the orthochronous, proper Lorentz group
$\tilde \tau: {\rm L}^\uparrow_+(d)\to {\rm Gl}(L)$, locality, spectral property and
cluster property.  Here $\tilde \tau$ is  obtained by analytic
 continuation of $\tau$
to a representation of the proper complex Lorentz group over $\, \C^d$
 (which contains $SO(d)$
as a real submanifold) and restriction of this representation to the
real orthochronous proper Lorentz group.
\end{theorem}
\noindent{\bf Proof.} That $S_n^T$ has a Fourier-Laplace representation with $\hat W_n^T$ defined as above
for the case $n\geq 3$ follows from the related representation for $S_{n,m_1,\ldots,m_n}^T$, the linearity
of the Fourier-Laplace transform and the general formula for differentiation of a 
Fourier-Laplace transform. That $W_n^T$ is tempered can be derived as in \cite{AGW2} or \cite{AG}. The formula for $n=2$ can be derived as in the case of the two-point
function of the free field. The properties of $W_n^T$ now follow from the general formalism of
analytic continuation, \cite{OS1}\kasten 

\section{Quantum fields with indefinite metric}
In this section we show that the Wightman functions constructed in Theorem \ref{3.1theo}
can be considered as vacuum expectation values of some quantum field theory (QFT) with 
indefinite metric \cite{MS}.

We first introduce the concept of a QFT with indefinite metric: Let ${\cal H}$ be a (separable)
Hilbert space and ${\cal D}\subseteq {\cal H}$ a dense domain. Let $\eta$ be a self-adjoint operator on ${\cal H}$
s.t. $\eta^2=1$. $\eta$ is called the \underline{metric operator}. We define ${\sf O}_\eta({\cal D})$ as the unital, involutive algebra of (unbounded) Hilbert space operators
$A:{\cal D}\to{\cal D}$ s.t. $A^{[*]}=\eta A^*\eta|_{\cal D}:{\cal D}\to{\cal D}$ exists. Here $A^*$ stands for the  Hilbert space adjoint of $(A,{\cal D})$. The canonical topology on ${\sf O}_\eta({\cal D})$ is induced by the seminorms $A\to|(\Psi_1,\eta A\Psi_2)|$, $\Psi_1,\Psi_2\in{\cal D}$. 

We say that a sequence of Wightman functions $\{ W_n \}_{n\in\N_0}$, $W_n\in{\cal S}_n'$ (where ${\cal S}_n={\cal S}^{\otimes n}$ and $'$ stands for the topological dual - in contrast to the previous sections test functions and distributions from now on are complex valued ) fulfills the \underline{Hilbert space structure condition (HSSC)} if on ${\cal S}_n$ there is a Hilbert seminorm $p_n$ s.t. 
\begin{equation}
\label{4.1eqa}
|W_{j+l}(f\otimes h)|\leq p_j(f)p_l(h)~~~\forall f\in {\cal S}_j,h\in{\cal S}_l, l,j\in\N.
\end{equation}
 We note that the HSSC for $\{W_n\}_{n\in\N_0}$ is implied
by the existence of a Schwartz norm $\|.\|$ on ${\cal S}$ s.t. $\hat W_n^T\in{\cal S}_n'$ is a continuous distribution w.r.t. $\|.\|^{\otimes n}$ on ${\cal S}_n$ (since $\cal S$ is a nuclear space, the tensor product of norms is well-defined) \cite{AGW2,Ho1}.

\begin{theorem}
\label{4.1theo}
Let $\{W_n\}_{n\in\N_0}$ be a sequence of Wightman functions which fulfill the requirements of temperedness, $\tilde\tau$-covariance, spectrality, locality and Hermiticity. If furthermore the HSSC (\ref{4.1eqa}) holds, then there exists an algebra ${\sf O}_\eta({\cal D})$ acting on a separable Hilbert space ${\cal H}$ with a distinguished 
normalized vector $\Psi_0\in{\cal H}$, $\eta\Psi_0=\Psi_0$, ('vacuum'), an operator valued distribution $\phi:{\cal S}\ni f\to\phi(f)\in{\sf O}_\eta({\cal D})$ and a $\eta$-unitary continuous representation of the orthochronous, proper Poincar\'e group\footnote{${\sf P}^\uparrow_*(d)$ is the semidirect product of ${\sf L}^\uparrow_+(d)$ with the translation group $\R^d$.}  ${\sf U}:{\rm P}^\uparrow_+(d)\to {\sf O}_\eta({\cal D})$ (${\sf U}^{[*]}={\sf U}^{-1}$) such that

\noindent (i) ${\cal D}$ is generated by repeated application of operators $\phi(f), f\in{\cal S}$ on $\Psi_0$, $\Psi_0$ is invariant under the representation ${\sf U}$ of ${\rm P}^\uparrow_+$ and ${\sf U}$ fulfills the spectral condition $\int_{\R^d}(\eta\Psi_1,{\sf U}(a)\Psi_2)e^{ip\cdot a}\, da=0$ if $p$ is not in the forward lightcone (here $\cdot$ is the Minkowski inner product);

\noindent (ii)  $\phi$ is Hermitean $\phi^{[*]}(f)=\phi(f^*)$ where $f^*$ is the (component wise) complex conjugation of $f$; $\phi$ is local ($\phi(f)$ and $\phi(h)$ commute on ${\cal D}$ if the support of $f$ and $h$ are space-like separated: $(x-y)^2<0$ for $x\in{\rm supp}f,y\in{\rm supp} h$); $\phi$ transforms $\tilde\tau$-covariantly (${\sf U}(g)\phi(f){\sf U}(g^{-1})=\phi(f_{g^{-1}})$ $\forall g\in{\rm P}^\uparrow_+$ with $f_g(x)=\tilde\tau(g)f(g^{-1}x)$);

\noindent (iii) $W_n(f_1\otimes\cdots\otimes f_n)=(\Psi_0,\phi(f_1)\cdots\phi(f_n)\Psi_0)$ $\forall n\in\N, f_l\in{\cal S}$.
\end{theorem}
For the proof, a kind of GNS-construction on an inner product space is performed, see e.g. \cite{Sc,MS,Ho2}. It should be noted that the above assignment of a QFT with indefinite metric to a sequence of Wightman functions in general depends on the Hilbert seminorms $p_n$ in (\ref{4.1eqa}) and therefore is not 'intrinsic' for the sequence of Wightman functions. For an example see \cite{Ar}.

Concerning our models in Theorem \ref{3.1theo} we now get:

\begin{theorem}
\label{4.2theo}
The Wightman functions defined in Section 3 fulfill the HSSC (\ref{4.1eqa}). In particular, there exists a QFT with indefinite metric (cf. Theorem \ref{4.1theo}) s.t.
the Wightman functions are given as the vacuum expectation values of that QFT.
\end{theorem}
{\bf Proof.} By Theorem \ref{4.1theo} the Wightman functions fulfill all the requirements of Theorem 
\ref{4.1theo}, except for the HSSC. That also the HSSC holds, can be seen most easily by 
verifying a uniform continuity property w.r.t. $\|.\|^{\otimes n}$ for the truncated
 Wightman functions $\hat W_n^T$, as explained above. Here $\|.\|$ is some Schwartz 
 norm on ${\cal S}$. It has been verified in \cite{AG,AGW3} that there is such a uniform
  continuity for $\hat W_{n,m_1,\ldots,m_n}^T$ and thus also the linear combinations
   of these distributions in (\ref{3.7eqa}) have this property. But the Fourier transformed
    Wightman functions of our model are given by the multiplication of the described linear
	 combination by a polynomial $Q_n^M(k_1,\ldots,k_n)$ and it is thus sufficient to verify that the degree of $Q_n^M$ in any variable $k_l$ is bounded independently of $n$, since we then can replace the Schwartz norm $\|.\|$ by the Schwartz norm $\|(1+|k|^2)^{l/2}.\|$ for $l$ larger or equal to this uniform degree. That such a uniform bound of the degree
in the $k_l$ exists is a straight forward consequence of the definitions (\ref{2.11eqa}) and (\ref{3.5aeqa}).\kasten  

\section{On the construction of asymptotic states and the $S$-matrix}

Here we describe the scattering behavior of the QFT models with
 indefinite metric constructed in Theorem \ref{4.2theo}. Since the
  standard axiomatic scattering theory \cite{Ha,He,Ru,RS} heavily relies
   on positivity of the Wightman functions, we can not apply these methods here. 
   Let us therefore first consider the general problem of the construction of asymptotic
    states for quantum fields with indefinite metric following \cite{AG,Go1}:

Let ${\cal S}^{\rm ext}={\cal S}\otimes \C^3$ be the "extended test function space".
 We want to construct an (in general non-local) "extended quantum field" $\Phi:{\cal S}^{\rm ext}\to {\sf O}_\eta({\cal D})$, where the three components of $\Phi$ can be interpreted as the incoming- local- and outgoing field $\Phi=(\phi^{\rm in},\phi,\phi^{\rm out})$, using the GNS-like construction of Therorem \ref{4.1theo}. We define $J(f^{\rm in},f^{\rm loc},f^{\rm out})=f^{\rm in}+f^{\rm loc}+f^{\rm out}$ and furthermore $J^{\rm in/loc/out}:{\cal S}\to{\cal S}^{\rm ext}$ as the injection in the first/second/third component. Let the mass content of the theory be given by the masses $m_1,\ldots,m_N>0$, which in our case are determined by the representation (\ref{2.4eqa}) of the partial differential operator $D$. Let $\varphi\in C_0^{\infty}(\R,\R)$ with support in $(-\epsilon,\epsilon)$ such that $0<\epsilon<{\rm min}\{ m_l^2, |m_l^2-m_j^2|,l,j=1,\ldots,N, l\not=j\}$. We define $\chi^{\pm}(k,m)=\theta(\pm k^0)\varphi(k^2-m^2)$ and we set 
\begin{equation}
\label{5.1eqa}
\chi_t(a,k)=\left\{\begin{array}{ll}\sum_{l=1}^N\left[\chi^+(k,m_l)e^{-i(k^0-\omega_l)t}+ \chi^-(k,m_l)e^{-i(k^0+\omega_l)t}\right]&{\rm for ~} a={\rm in}\\
1&{\rm for ~} a={\rm loc}\\\sum_{l=1}^N\left[\chi^+(k,m_l)e^{i(k^0-\omega_l)t}+\chi^-(k,m_l)e^{i(k^0+\omega_l)t
}\right]&{\rm for~}a={\rm out}\end{array}\right.
\end{equation}  
Here $\omega_l=(|\vec k|^2+m^2_l)^{1/2}$. We now define $\Omega_t{\cal S}^{\rm ext}\to{\cal S}$ and $\Omega^{\rm in/out}:{\cal S}\to{\cal S}$ by
\begin{equation}
\label{5.2eqa}
{\cal F}\Omega^{\rm ext}\bar {\cal F}=\left(\begin{array}{ccc}\chi_t(k,{\rm in})&0&0\\
0&\chi_t(k,{\rm loc})&0\\
0&0&\chi_t(k,{\rm out})\end{array}\right)
\end{equation}
ant $\Omega_t=J\circ \Omega^{\rm ext}_t$, $\Omega^{\rm in/out}=\Omega_t\circ J^{\rm in/out}$. Here ${\cal F}$ ($\bar{\cal F}$) denotes the (inverse) Fourier transform. For $f_l\in{\cal S}^{\rm ext}$ we set
\begin{equation}
\label{5.3eqa}
F_n(f_1\otimes\cdots\otimes f_n)=\lim_{t_1,\ldots,t_n\to+\infty}W_n(\Omega_{t_1}f_1\otimes\cdots\otimes\Omega_{t_n}f_n
)
\end{equation}
provided that the limit exists and $F_n$ is in ${\cal S}^{\rm ext '}_n=({\cal S}^{{\rm ext}
\otimes n})'$. Here the limit $t_1,\ldots,t_n\to+\infty$ has to be understood in the
 sense that first one $t_l$ goes to infinity, then the next etc. and that the limit does
  not depend on the chosen order. If $F_n$  exists for any $n$ then the sequence
   $\{F_n\}_{n\in\N_0}$ is called the \underline{form factor functional}. 
   Heuristically, $\{F_n\}_{n\in\N_0}$ contains the 'mixed' vacuum expectation values of 
   in- loc- and out-fields. If the form factor functional fulfills the HSSC, then one can
    construct in-, loc- and out-fields acting on one Hilbert space $\cal H$
	 by applying the GNS-like construction of Theorem \ref{4.1theo} in order to 
	 construct a "extended quantum field" $\Phi$. Given $\Phi$, we then define
	  $\phi^{\rm in/loc/out}=\Phi\circ J^{\rm in/loc/out}$ \cite{AG}. The main result
	   for the present section is that this whole construction is possible for the models defined in Therorem \ref{4.2theo}. To formulate this, we require some  more definitions: We set
\begin{equation}
\label{5.4eqa}
\hat \Delta_{m} (a,k)=\left\{
\begin{array}{ll}
 -i\pi (\delta_{m}^+(k)-\delta_{m}^-(k)) & \mbox{ for $a=$in}\\
1 \left/(k^2-m^2)\right.                & \mbox{ for $a=$loc}\\
i\pi (\delta_{m}^+(k)-\delta_{m}^-(k)) & \mbox{ for $a=$out}
\end{array}\right.
\end{equation}
and we define the Fourier-transformed truncated form factor functional associated to the 
Wightman functions $W_{n,m_1,\ldots,m_n}^T$ defined in Section 3 for $n\geq3$ $a_l=$in/loc/out via 
\begin{eqnarray}
\label{5.5eqa} 
&&\hat F_{n,m_1,\ldots,m_n}^{T(a_1,\ldots,a_n)}(k_1,\ldots,k_n)\nonumber\\
&=&-(2\pi)^{-((dn-2)-2)/2}\left\{\sum_{j=1}^n\prod_{l=1}^{j-1} \delta^-_{m_l}(k_l) \hat
\Delta_{m_j}(a_j,k_j)\prod_{l=j+1}^n\delta^+_{m_l}(k_l)\right\}\delta
(\sum_{l=1}^nk_l)\nonumber \\
\end{eqnarray}
and we finally define the truncated, Fourier transformed form factor functional for our models as $F^{T(a_1,a_2)}_2=W^T_2$, $a_1,a_2=$in/loc/out and for $n\geq3$ we set in analogy to Eq. (\ref{3.7eqa})
\begin{eqnarray}
\label{5.6eqa}
\hat F_{n,\alpha_1\cdots\alpha_n}^{T(a_1,\ldots,a_n)}(k_1,\ldots,k_n)&=&Q^M_{n,\alpha_1,\ldots,\alpha_n}( k_1,\ldots,
 k_n)\nonumber\\
 &\times&\sum_{l_1,\ldots,l_n=1}^N\prod_{j=1}^nb_{l_j} \hat F_{n,m_{l_1},\ldots,m_{l_n}}^{T(a_1,\ldots,a_n)}(k_1,\ldots,k_n)
\end{eqnarray}
We define $\{F_n\}_{n\in\N_0}$ to be the sequence of distributions associated with the truncated sequence $\{F_n^T\}_{n\in\N}$ defined above. We get
\begin{theorem}
\label{5.1theo}
Let $\{W_n\}_{n\in\N_0}$ be the Wightman functions of the QFT-models defined in Theorem \ref{4.2theo}. Then the associated form factor functional is given by $\{F_n\}_{n\in\N_0}$ and fulfills the HSSC. Thus, 

\noindent (i) There exists a QFT with indefinite metric (over ${\cal S}^{\rm ext}$) (${\cal H},\eta,\Psi_0,\Phi,{\sf U})$ fulfilling Hermiticity, spectrality and clustering  s.t. the fields $\phi^{\rm in/loc/out}=\Phi\circ J^{\rm in/loc out}$ in addition are $\tilde\tau$-covariant and local.

\noindent (ii) The asymptotic fields $\phi^{\rm in/out}$ are given as a sum of independent free $\tilde\tau$-vector fields fields with masses $m_1,\ldots,m_N$. 

\noindent (iii) $\phi=\phi^{\rm loc}$ fulfills the LSZ asymptotic condition \cite{LSZ} w.r.t. $\phi^{\rm in/out}$, i.e.
\begin{equation}
\label{5.7eqa}
\lim_{t\to+\infty}\phi(\Omega^{\rm in/out}_tf)=\phi^{\rm in/out}(f)~~\forall f\in{\cal S}
\end{equation} 
where the convergence is in ${\sf O}_\eta({\cal D})$.
\end{theorem}
{\bf Proof.} That $W_{n,m_1,\ldots,m_n}^T$ and $F_{n,m_1,\ldots,m_n}^T$ fulfill Eq.
 (\ref{5.3eqa}) (here truncation plays no r\^ole, cf. Proposition 3.4 of \cite{AG}),
  can be seen as follows: The relation has been proven for a single mass 
  $m$ in \cite{AG}. The proof can be extended by a simple adaptation of notation 
  for the case where $m_1,\ldots,m_n$ are different masses and the multipliers $\chi_t(k_l,a)$ depend 
  on only one mass depending on $l$. Given this observation, let us consider our more
   complicated multipliers in Eq. (\ref{5.1eqa}). 
   If a factor $\chi^{\pm}(k,m_l)e^{\pm i(k^0\pm\omega_l)t}$ is multiplied 
   with a factor $\delta^\pm_{m_j}(k)$, then the result is zero by the support
    properties of $\chi^\pm(k,m_l)$ whenever $j\not=l$. Likewise,
	 if a factor $1/(k_j^2-m_j^2)$ is being multiplied with such a factor, then this expression
	  vanishes in the limit $t\to\infty$ by the Riemann-Lebesgue lemma \cite{RS}. We thus see that only those terms with the "right" masses count in (\ref{5.1eqa}) and this establishes the result for $W_{n,m_1\ldots,m_n}^T$ and $F^{T(a_1,\ldots,a_n)}_{n,m_1,\ldots,m_n}$, $n\geq 3$. The case $n=2$ is trivial. 

The statement that $\{F_n\}$ is the form factor functional associated to $\{W_n\}$ now follows from the fact that multiplication by $Q_{n}^M$ in energy-momentum space and the limit in (\ref{5.3eqa}) can be interchanged.

That the so-defined form factor functional fulfills the HSSC can be proven in a similar way as in Theorem \ref{4.2theo}, cf. \cite{AG}. The remaining statements of the theorem then follow by the GNS-like construction as in Theorem \ref{4.1theo}, cf. \cite{AG} for the details. \kasten 

It should be remarked that the metric on the asymptotic Hilbert spaces
 ${\cal H}^{\rm in/out}$ generated by repeated application of the fields 
 $\phi^{\rm in/out}$ to the vacuum
depends on the properties of $C^{\beta_1,\beta_2}$, $b_l$ and $Q_E$
 and gauge principles for the asymptotic fields have to be developed depending 
 on these quantities. For a single, scalar field ($C>0$, $b=1$, $Q_E=1$) these spaces carry a
  positive semi-definite metric \cite{AG}.

Finally we want to show that the scattering of the fields in Theorem \ref{5.1theo} is non-trivial. Given the form factor functional, one can define the $S$-matrix of the theory via
\begin{eqnarray}
\label{5.8eqa}
&&S_{r;n-r}(f_1\otimes\cdots\otimes f_r;f_{r+1}\otimes\cdots\otimes f_n)\nonumber\\
&=&F_n(J^{\rm in}f_1\otimes \cdots\otimes J^{\rm in}f_r\otimes J^{\rm out}f_{r+1}\otimes\cdots\otimes J^{\rm out}f_n)
\nonumber \\
&=&\left\langle \phi^{\rm in}(f_1^*)\cdots\phi^{\rm in}(f_r^*)\Psi_0,\phi^{\rm out}(f_{r+1})\cdots\phi^{\rm out}(f_n)\Psi_0\right\rangle
\end{eqnarray}
where $\langle.,.\rangle=(.,\eta.)$ is the indefinite inner product on ${\cal H}$.

Using the definitions (\ref{5.4eqa})-(\ref{5.6eqa}) one can
 verify by an explicit calculation the following corollary:

\begin{corollary}
\label{5.1cor}
The $S$-matrix of the models in Theorem \ref{5.1theo} is non-trivial (if $\psi(t)$ has a Poisson part). The Fourier transformed, truncated $S$-matrix is given by $S^T_2(k_1;k_2)=\hat W_2^T(k_1,k_2)$ and
\begin{eqnarray}
\label{5.9eqa}
&& \hat S_{r;n-r,\alpha_1\cdots\alpha_r;\alpha_{r+1}\cdots\alpha_n}^T(k_1,\ldots,k_r;k_{r+1}, \ldots,k_n)\nonumber \\
&=&-i(2\pi)^{-((dn-2)-4)/2}
Q^M_{\alpha_1,\ldots,\alpha_n}(k_1,\ldots,k_n) \sum_{l_1,\ldots,l_n=1}^N\prod_{j=1}^nb_{l_j}\nonumber\\
&\times&\prod_{j=1}^r \delta^{-}_{m_{l_j}}(k_l)\prod_{l=r+1}^n\delta^+_{m_{l_j}}(k_l)\, \delta(\sum_{l=1}^nk_l)
\end{eqnarray}
 for $n\geq 3$ where $k_1^0,\ldots,k_r^0<0$ and $k_{r+1}^0,\ldots,k_n^0>0$. 
 \end{corollary}

 We remark that in- and out- fields are free fields and fulfill canonical commutation 
 relations, the relations given in Corollary \ref{5.1cor} suffice
 to determine the whole scattering matrix.

\

\small
\noindent {\bf Acknowledgements.} We would like to thank C. Becker and R.
Gielerak for interesting discussions. The financial support of D.F.G.
via Project "Stochastische Analysis und Systeme mit unendlich vielen Freiheitsgraden"
 is gratefully acknowledged.

\end{document}